\documentclass[conference, letterpaper]{IEEEtran}
\usepackage[letterpaper,left=45pt,right=45pt,top=72pt,bottom=54pt]{geometry}

\usepackage{subcaption,graphicx}
\usepackage{xcolor}
\PassOptionsToPackage{hyphens}{url}\usepackage{hyperref}

\ifCLASSINFOpdf
\else
\fi
\usepackage{amsmath}

\usepackage{siunitx}
\usepackage{stfloats}
\usepackage{tikz}
\usepackage{textcomp}
\usepackage{hyperref}
\usepackage{lipsum}

\begin{document}
%
\title{Brain-based control of car infotainment}

\author{\IEEEauthorblockN{Andrea Bellotti$^{1,3}$,
Sergey Antopolskiy$^{2}$,
Anna Marchenkova$^{4}$,
Alessia Colucciello$^{2}$\\\\
Pietro Avanzini$^{4}$,
Giovanni Vecchiato$^{4}$,
Jonas Ambeck-Madsen$^{1}$,
Luca Ascari$^{2}$
} \\
\IEEEauthorblockA{$^{1}$ Toyota Motor Europe NV/SA, Hoge Wei 33B Zaventem, 1930, Belgium}
\IEEEauthorblockA{$^{2}$ Camlin Italy S.r.l., Strada Budellungo 2, Parma, 43123, Italy}
\IEEEauthorblockA{$^{3}$ Industrial Engineering Department, University of Parma, Parco Area delle Scienze 181/A, Parma, 43124, Italy}
\IEEEauthorblockA{$^{4}$ Institute of Neuroscience, National Research Council of Italy, Via Volturno 39/E, Parma, 43125, Italy}}


\newcommand\copyrighttext{%
  \footnotesize \textcopyright 2019 IEEE. Personal use of this material is permitted.
  Permission from IEEE must be obtained for all other uses, in any current or future
  media, including reprinting/republishing this material for advertising or promotional
  purposes, creating new collective works, for resale or redistribution to servers or
  lists, or reuse of any copyrighted component of this work in other works.
  DOI: \href{https://ieeexplore.ieee.org/document/8914448}{10.1109/SMC.2019.8914448}}
\newcommand\copyrightnotice{%
\begin{tikzpicture}[remember picture,overlay]
\node[anchor=south,yshift=10pt] at (current page.south) {\fbox{\parbox{\dimexpr\textwidth-\fboxsep-\fboxrule\relax}{\copyrighttext}}};
\end{tikzpicture}%
}

\maketitle

\copyrightnotice

\begin{abstract}
Nowadays, the possibility to run advanced AI on embedded systems allows natural interaction between humans and machines, especially in the automotive field.
We present a custom portable EEG-based Brain-Computer Interface (BCI) that exploits Event-Related Potentials (ERPs) induced with an oddball experimental paradigm to control the infotainment menu of a car. 
A preliminary evaluation of the system was performed on 10 participants in a standard laboratory setting and while driving on a closed private track. The task consisted of repeated presentations of 6 different menu icons in oddball fashion. Subject-specific models were trained with different machine learning approaches on cerebral data from either only laboratory or driving experiments (in-lab and in-car models) or a combination of the two (hybrid model) to classify EEG responses to target and non-target stimuli. All models were tested on the subjects' last in-car sessions that were not used for the training.
Analysis of ERPs amplitude showed statistically significant (p $<$ 0.05) differences between the EEG responses associated with target and non-target icons, both in the laboratory and while driving.
Classification Accuracy (CA) was above chance level for all subjects in all training configurations, with a deep CNN trained on the hybrid set achieving the highest scores (mean CA = 53 $\pm$ 12 \%, with 16 \% chance level for the 6-class discrimination). The ranking of the features importance provided by a classical BCI approach suggests an ERP-based discrimination between target and non-target responses. No statistical differences were observed between the CAs for the in-lab and in-car training sets, nor between the EEG responses in these conditions, indicating that the data collected in the standard laboratory setting could be readily used for a real driving application without a noticeable decrease in performance. However, refining the model with real-world data might be beneficial.
While there is still room for improvement, the results demonstrate the feasibility of a brain-based control of the car infotainment while driving.
\end{abstract}

%
\IEEEpeerreviewmaketitle

\section{Introduction}
The recent technological advances in mobile EEG systems concerning comfort, portability and reliability are stimulating researchers to design Brain-Computer Interfaces (BCI) for real-life conditions and perform experiments that take place outside the laboratory settings. With some joint effort this could eventually evolve into the hoped transition towards a reliable and wide-spread adoption of EEG-based assistance systems in the real world.\\
A number of studies have already achieved promising results confirming the ability to measure reliable EEG patterns, including  Event-Related Potentials (ERPs), outside the laboratory and while involved in real-life tasks.
For instance, a mobile EEG cap with dry electrodes has been successfully used to detect hazardous piloting maneuvers in real flight conditions \cite{scholl2016classification}.
An active wet EEG system has been employed to characterize Auditory Evoked Potentials while outdoor cycling, confirming similar ERP topography and morphology (with just slightly lower amplitudes), despite the introduced noise, in comparison with the controlled laboratory conditions \cite{scanlon2017taking}.
Other studies have, instead, adopted the EEG technology in real-world driving conditions to detect emergency breaking intentions \cite{haufe2014electrophysiology} or mismatches between the steering actuations and the drivers' intentions \cite{zhang2015eeg}.\\
Considering these automotive applications of BCIs it is clear how the human-vehicle interaction could be greatly enhanced employing portable infrastructures able to reliably interpret the drivers' intentions or reactions while driving.
Accordingly, there is an increasing number of car manufacturers working on more context-aware driving assistants to be integrated in their ADAS (Advanced Driving Assistance System), for instance for drowsiness detection or lane change assistance \cite{SILVEIRA2015}, \cite{6611147}, \cite{7379680}.\\
In line with the trend, we did a step forward in bringing BCIs in the real world by developing an easy-to-wear wireless EEG-based BCI to control the infotainment menu of a car. It is based on the presentation of the menu icons (6 in our implementation) one after another following a version of an oddball paradigm, and relies on the discrimination of the Event-Related Potentials (ERPs) elicited by the appearance of the desired (target) menu item from those associated with the other (non-target) icons.
A list of subject-specific classification models was trained with different machine learning approaches: two Convolutional Neural Networks presented in \cite{schirrmeister2017deep}, a custom CNN in between the two just mentioned, and a classical BCI approach based on manual feature extraction \cite{7296414}, \cite{ABOOTALEBI200948} and a random forest classifier. In the present paper we describe the performance of this BCI evaluated on the data collected from a group of 10 subjects in a controlled laboratory environment and in real-life driving conditions.
\newgeometry{top=54pt, bottom=54pt, left=45pt,right=45pt}

\section{Materials and methods}
\subsection{Participants}
Ten subjects (9 male and 1 female, age 37.8 $\pm$ 7.42) participated in the experiments in both laboratory and driving scenarios. None of the involved subjects were professional drivers, but all had valid driving licenses and driving experience. All subjects had normal or corrected-to-normal vision, did not report any known neurological or psychiatric diseases at the time of the experiments and provided written informed consent for participation. 

\subsection{Acquisition system}
\begin{figure}[!h]
    \centering
    \includegraphics[width=0.9\linewidth]{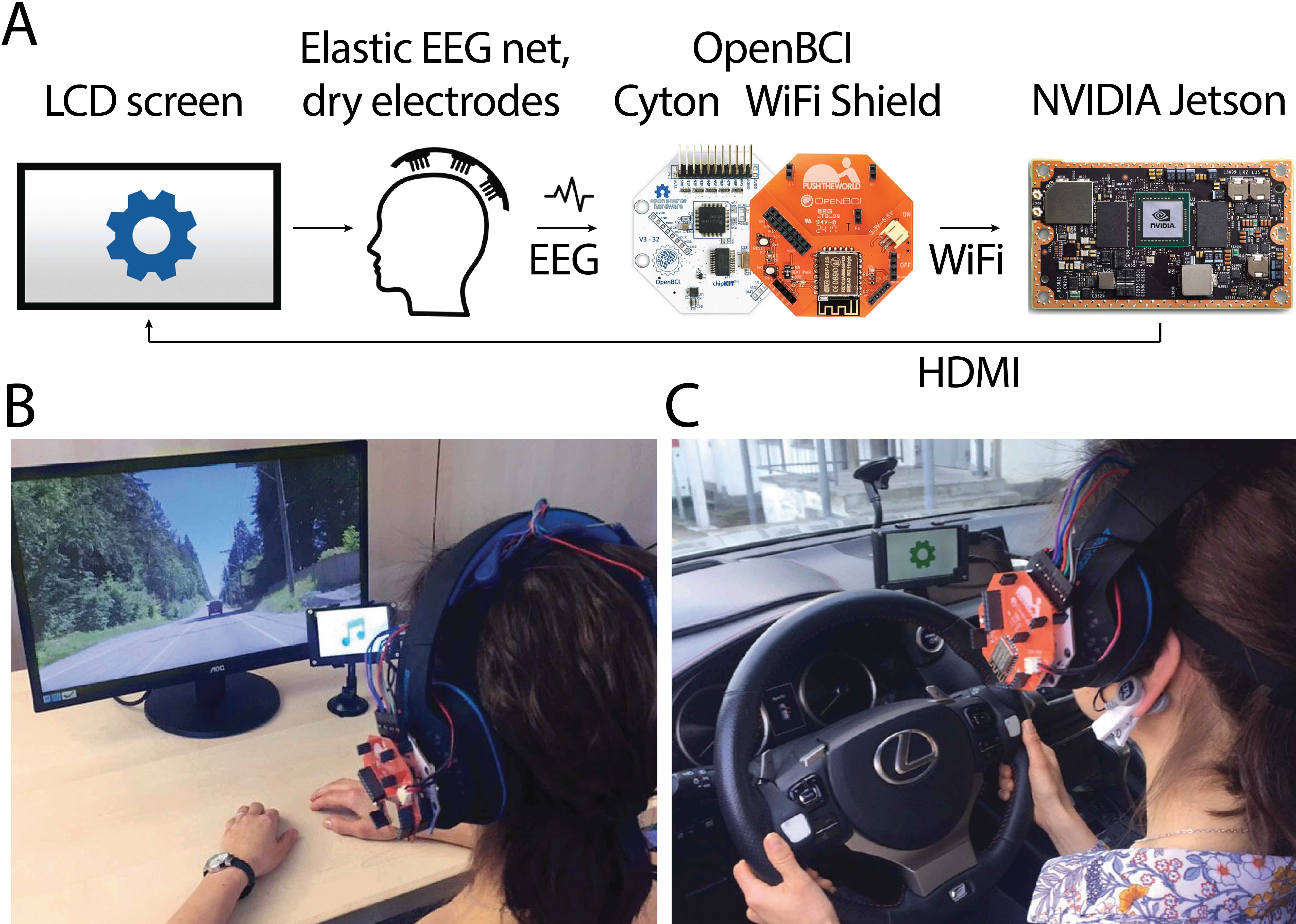}
    \caption{A, Schematic representation of the acquisition system. B, In-lab experimental setup. C, In-car experimental setup.}
    \label{fig:Recording_system_setup}
 \end{figure}

Experiments were performed using a portable wireless EEG recording system made up by the following commercially available components: dry Neuroelectrics EEG electrodes, custom-designed elastic EEG cap, OpenBCI Cyton Board with OpenBCI Wifi Shield, NVIDIA Jetson TX2 Module. The schematic representation of the setup is shown on the Fig. \ref{fig:Recording_system_setup}, panel A.
\\
The elastic EEG cap prototype was designed with the main goal of being straightforward and effortless to use but at the same time providing good-quality EEG recordings. For each cap we used Neuroelectrics dry electrodes that were held in positions Cz, Pz and Fp1 by a net of adjustable elastic bands, providing an ergonomic head fit and adequate pressure to ensure a good connection with the skin, while being comfortable to wear for long stretches of time. The Neuroelectrics EarClip dual electrode was used for reference and ground on the left earlobe.
OpenBCI was used for collecting raw EEG signals from the EEG electrodes (sampling rate = 500 Hz) and for transmitting them via WiFi to the Jetson, which was responsible for creating the corresponding LSL (Lab Streaming Layer) stream.
The Jetson also created additional streams carrying information about lost packets and temporary connection loss. All streams were saved as an XDF file by the recording program included in the library (LabRecorder). The OpenBCI board and its battery were mounted on a set of standard, commercially available headphones worn on top of the EEG net. This assembly mechanically decoupled the electrodes from the recording system, thus improving electrode stability throughout the experiment lowering the amount of noise introduced by the head movements and the movement of the car. The Jetson was powered by a portable battery and used for storing the data, running the stimuli presentation software and calculating the online predictions in the in-car scenario.\\
During both in-lab and in-car experiments, visual stimuli were presented on a 5-inch LCD touch screen with 800 x 480 resolution (later referred to as "small screen"), connected to the Jetson via HDMI cable, Fig. \ref{fig:Recording_system_setup}, panel A. The script for presenting the stimuli was written using the Python library PsychoPy, which can generate precise and reliable visual stimuli with negligible delays \cite{garaizar2014accuracy}. In a separate set of experiments we tested the system's synchronization, comparing the timing of the LSL markers on the EEG stream with the actual appearance of the stimuli on the screen determined by the photoresistor \cite{garaizar2014accuracy}. The resulted delay was 30 $\pm$ 2.7 ms, with a very narrow distribution, that remained consistent between the recordings.

\subsection{Experimental settings in the laboratory}
\begin{figure}[!t]
    \centering
    \includegraphics[width=0.99\linewidth]{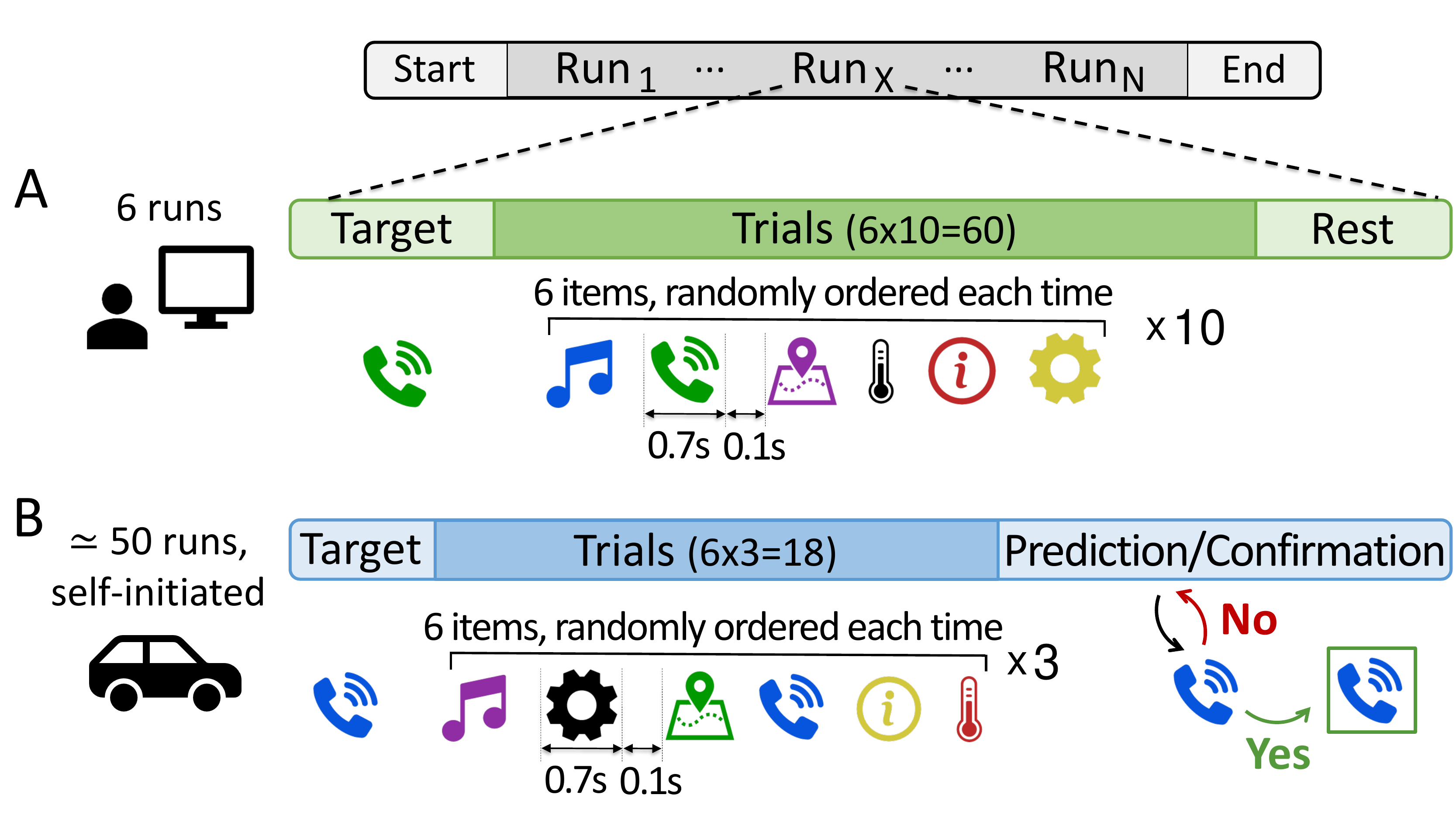}
    \caption{Experimental session design. A, In-lab condition: one session comprised 6 runs (each having 60 trials; each icon repeating 10 times) of visual oddball task, with 30 s rest periods in between. B, In-car condition: one session included around 50 runs (each having 18 trials; each icon repeating 3 times) that were self-initiated by the subject. In both conditions the colors for the items were randomly chosen in the beginning of the session and remained constant within the session; items in a run were block-randomized on each repetition of a 6-item block.}
    \label{fig:Experimental_Protocol_design}
 \end{figure}

The first set of experiments was performed in a standard laboratory environment. Participants were seated 1 m in front of a 24-inch computer monitor with 1920 x 1080 resolution and completed a visual oddball-like task similar to \cite{luck1994electrophysiological} and  \cite{hoffmann2008efficient}, mentally focusing on one randomly-chosen icon (target) from a set of 6 icons, that were appearing one by one in a random order for 700 ms with the inter-stimulus interval of 100 ms. Icons represented possible menu items in a car navigation menu and were shown on a small screen positioned in front of the monitor, on the periphery of the subject's visual field (to simulate in-car dashboard position). Meanwhile, the computer monitor showed a driving video recorded from a first-person perspective (to simulate driving condition) as shown in Fig. \ref{fig:Recording_system_setup}, panel B. Subjects were requested not to look at the small screen and the icons directly, but focus on them with their peripheral vision, keeping the eyes in the central part of the computer monitor; they were asked to refrain from moving the head and the body, excessive blinking and excessive eye movements.\\
The schematic representation of the in-lab session layout is shown in Fig.  \ref{fig:Experimental_Protocol_design}, A. Each experimental session consisted of 6 runs of the visual oddball task, with 30 s breaks between runs. In each run, the icons were flashed 60 times: each of the 6 icons was presented 10 times in block-randomized order, so within every 6 presentations each icon appeared 1 time. This to allow us to later compute the optimal number of repetitions that would be actually used in real driving conditions based on the tradeoff between performance and time needed to get the feedback. Icons were of the same size (600 x 600 pixel), but had different shapes and colors for better discrimination \cite{liu2010comparison}. Colors were randomly selected among 6 colors - black, green, yellow, magenta, red, blue - at the beginning of every session and were kept constant for all the runs in the session.\\
All ten subjects completed 6 in-lab sessions divided between 2 days (with the interval of 4.25 $\pm$ 2.33 days). Both subject s001 and s005 completed one additional day of in-lab recordings (3 sessions) because of signal quality issues and/or poor wireless connection in one of the first recording days. \\
Based on the collected data, a classification model was built for each subject to discriminate on a single-trial basis if a particular trial was target or non-target. For the details on the classification model see the corresponding section below.

\subsection{Experimental settings on the proving ground}
The real car experiments were performed by the same participants while driving a Toyota Prius in controlled conditions on Toyota's closed private proving ground (1.2 Km flat oval track). During the experiments, the drivers kept the speed between 45 and 55 Km/h and activated the system (initiated the runs) autonomously when ready.\\
The stimuli for the oddball task were presented on the same small screen used for the laboratory experiments, placed on the dashboard in front of the subject behind the steering wheel, as on the Fig. \ref{fig:Recording_system_setup}, C. The driving experiments consisted of 3 in-car sessions for each subject, each session was completed on a separate day with the average inter-session interval of 3.5 $\pm$ 5.86 days. The session layout was modified to avoid long periods of heavy mental load and allow the user to self-initiate every trial when ready and not distracted as it normally happens when manually interacting with the infortainment of the car. The outline of the in-car session layout is shown in Fig. \ref{fig:Experimental_Protocol_design}, B. Each in-car session lasted approximately 1 hour, including preparation time; during the experiment participants were free to self-initiate each run by pressing a button on the steering wheel when ready. At this preliminary stage of the study we used a button to activate the system, instead of another advanced hand-free means, because we wanted to focus on the feasibility of the brain-controlled selection of the items of the car infotainment menu while driving rather than building a final product and thus all other details were left as simple and automatic as possible. Additionally, runs were shortened compared to the laboratory settings and comprised 18 trials (3 presentations for each icon, block-randomized as in in-lab experiments). During the run the EEG data were continuously buffered and then classified by simple subject-specific models trained on all the previously recorded in-lab data of the subject to provide him/her online feedback. This was done primarily to increase the engagement and attention of the subjects during the task and, additionally, for the initial evaluation of the classification model, trained on the data collected during the in-lab experiments. 

\subsection{Data acquisition and analysis}
To make the system easy-to-use and portable, EEG signals were recorded from 3 electrode locations (according to the 10/20 system): Cz, Pz, Fp1 and referenced to the left earlobe. The choice for the electrode positions was made in accordance with the central-parietal distribution of the P300 ERP response to rare target stimuli \cite{kirchner2018multi}, which was expected as a major outcome of the oddball experiments; channel Fp1 was used for ocular artifacts detection. The raw EEG signals were acquired by the OpenBCI with a sampling rate of 500 Hz and saved as LSL streams along with the experimental event markers. EEG signals were then band-pass filtered between 0.1 and 30 Hz to increase reliability and signal to noise ratio while preserving the relevant physiological information \cite{scanlon2017taking, kuziek2017transitioning}. EEG data for each trial were extracted using Python MNE library in the time window [-100:700] ms around the onset of experimental events for all recorded channels, using the first 100 ms (from -100 ms to stimulus onset set to 0) for baseline correction.\\
Rejection of bad trials was done automatically in two steps. Due to the imperfections in wireless data transmission, inevitable minor data loss occurred in some trials, which were removed from the analysis if more than 20 consecutive samples got missing. As step 2, a threshold-based artifact rejection algorithm was implemented to remove from the subsequent analysis those trials having signals exceeding the [-100, 100] $\mu$V threshold in any of the analyzed channels (Cz, Pz, Fp1). At this step, the rejection rate was calculated as percentage from the epochs remaining from the step 1 rejection. Total rejection rates were calculated as percentage of all rejected epochs with respect to the total number of epochs for each subject, and then averaged across subjects. 
After pre-processing, the extracted trials were used to train the classification models and performance evaluation, and for the analysis of the ERPs to target and non-target stimuli. For the latter, we analyzed the extracted EEG signals in the time window from 250 to 530 ms after the stimulus onset for each trial and calculated peak ERP amplitude (maximum in the window), peak-to-peak amplitude (the difference between the minimum and the maximum) and ERP peak latency (in ms from stimulus onset) for each trial of channels Cz and Pz. The window was chosen in order to include the P300 ERP component, that appears in response to the target stimuli in the oddball paradigm, with a peak latency that could vary depending on the exact experimental protocol \cite{luck1994electrophysiological, hoffmann2008efficient}. Data of channels Cz and Pz were averaged with the median operator for each subject and then among subjects to obtain grand averages. The differences between target and non-target responses, and between target responses in different training sets were assessed using a pairwise comparison with Welch's t-test for unequal variances with a Bonferroni correction for multiple comparisons. The average number of trials per subject that were analyzed: 1930 $\pm$ 448 for in-lab, and 2385 $\pm$ 153 for in-car experiments, with the 1 to 5 target/non-target ratio.
To estimate the amount of noise on single trials, we calculated root mean square (RMS) scores, randomly selecting 200 trials for each subject from in-lab and in-car experiments, performing 1000 permutations. RMS scores were averaged first for channels (Cz and Pz), than for subjects, and then the grand averages between subjects was computed for each condition \cite{kuziek2017transitioning}. Power spectral density (PSD) was calculated for all EEG trials segments in the [0.1, 30] Hz window for channels Pz and Cz for each participant, using Welch method (python scipy package; with 256 samples segments, 200 sample overlap and 1024 points Fast Fourier Transform); the results were then averaged between subjects with the median operator to compute the grand averages for both in-lab and in-car conditions.

 \begin{figure*}[!t]
    \centering
    \includegraphics[width=0.8\linewidth]{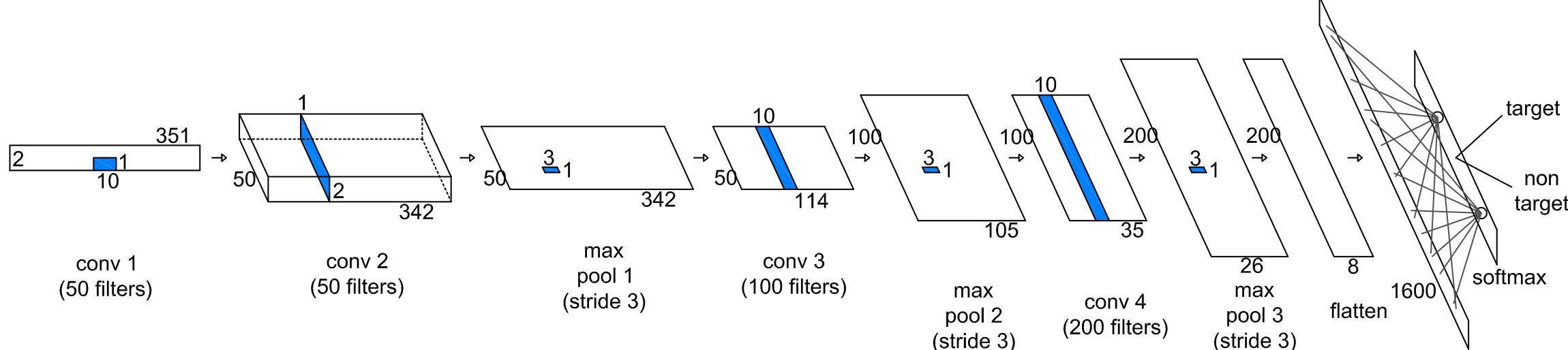}
    \caption{Architecture of the custom intermediate CNN. Blue shapes represent the convolution/pooling kernels.}
    \label{fig:cnn}
 \end{figure*}

\subsection{Classification models}
\begin{figure}[!b]
    \centering
    \includegraphics[width=0.99\linewidth]{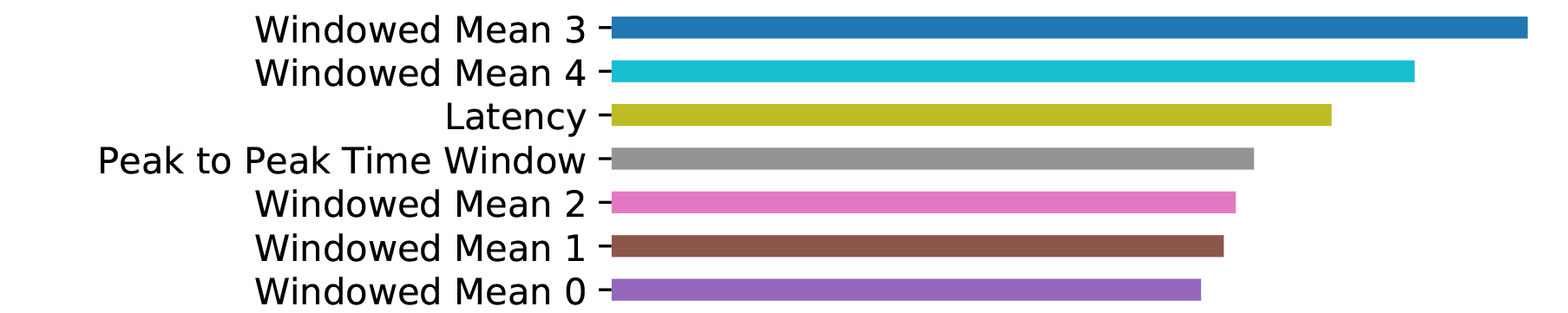}
    \caption{Ranking of the most important subset of features provided by the random forest classifier, averaged across training sets and subjects.}
    \label{fig:features}
 \end{figure}

To build the subject-specific classification models we took inspiration from the convolutional architectures presented in \cite{schirrmeister2017deep}, which demonstrated to be effective in a Motor Imagery (MI) classification task outperforming the golden standard FBCSP (Filter Bank Common Spatial Pattern) approach. Being very flexible architectures performing classical time/space convolutions, we hypothesized that these CNNs would work also for an ERP-related task. We trained both the original deep and shallow networks adopting the default parameters presented in the paper and implemented a custom intermediate CNN to reduce computational time while trying to improve the classification performance by reducing the possible overfitting due to the large number of parameters of the deep network. We changed several architectural details to better fit our experimental settings and classification goal (Fig. \ref{fig:cnn}). First, performing cross-validation we found that for the custom intermediate network, using 50 filters instead of 25 in both convolutional layers of Conv-Pool Block 1 and adopting a Relu activation function instead of an Elu resulted in a higher classification accuracy; second, we entirely dropped the Conv-Pool Block 2 to reduce the number of parameters of the network; third, we decreased the dropout probability rate from 0.5 to 0.1 in all the network blocks again based on the cross-validation results. Lastly, we modified the classification layer to produce a single-trial binary classification (target vs non-target). These modifications halved the training time from 25 s to 17 s per training epoch.
For each machine learning approach we used the same architectural and learning parameters for all the subjects. With the deep and shallow networks we used the default learning parameters of the Braindecode library whereas with our custom intermediate network we adopted Stochastic Gradient Descent (SGD) with Nestorov momentum (momentum $\mu=0.95$, learning rate $\alpha=1e^{-5}$, weight decay $\lambda=1e^{-6}$) as optimizer instead of Adam keeping as loss function the categorical cross-entropy, and applied batch normalization considering batches of size 32 instead of 64. All these choices were the result of a grid search we implemented to further improve the classification accuracy. \\
To compare the performances obtained with the CNN approaches with a more classical feature-based classification approach, we used random forest classifier with a combination of features extracted with windowed means paradigm presented in \cite{7296414} and the morphological features presented in \cite{ABOOTALEBI200948}. In particular, we extracted 5 windows in the time interval between 200 ms and 600 ms and computed the features of channels Cz and Pz, obtaining 44 features in total. Based on the results of cross-validation, the depth and number of trees for the random forest were set to 12 and 1000 respectively. Fig. \ref{fig:features} shows the ranked importance of the first features provided by the random forest classifier demonstrating that P300-related features like positive peak latency, peak-to-peak time window and windowed means around the positive peak were the most relevant for the classification.
For all the evaluated approaches, to account for the imbalance in the dataset (1/5 ratio of target and non-target trials), we forced a class re-weighting in the fitting phase by applying a five-fold increase to the weight of the target trials. \\
For model fitting and evaluation we used EEG segments from channels Cz and Pz, in the [0:700] ms window from stimulus onset. 
The prediction of the correct icon was computed by grouping the 18 trials (3 repetitions of each presented stimulus) of a run based on the icon, averaging the single-trial probabilities of each icon group separately and providing the highest as output.
To find the optimal subject-specific number of training epochs for each individual deep learning approach, we implemented a 5-fold stratified cross-validation procedure. With the deep and shallow networks we directly used as metric the averaged balanced classification accuracy obtained on the validation set, whereas with the custom intermediate network, where we had more control on the training phase, we simulated the aggregation procedure by performing the same procedure on 100 permutations of the validation set subdivided into stratified blocks of 18 trials (3 repetitions) representing single runs. This allowed us to obtain a more accurate tuning of hyperparameters. \\
Once we identified the optimal hyperparameters (number of training epochs for deep learning approaches, number of windowed means, depth and number of trees for the classical approach), we re-trained on all training set data and evaluated the performance of the models on the held-out test set. Since we were mostly interested in applying and evaluating the models in a real-life driving conditions, we saved each subjects' last in-car session as a training set (average number of trials for each subject was 881 $\pm$ 58). \\  
To evaluate the differences between the data from the in-lab and in-car sessions, and how these different recording modes could affect the final classification scores, we repeated the training procedure with three different training sets. The first one consisted of all the in-lab trials for each subject (1930 $\pm$ 448 trials), the second one included only first two in-car sessions (1581 $\pm$ 143 trials), and the third training set was the combination of the other two (3512 $\pm$ 457 trials). To account for the imbalance of trials of the hybrid training set with respect to the other two, we performed the training of each different machine learning approach 5 times sampling from the hybrid set only half of the shuffled trials in a stratified manner (thus preserving the proportions of the classes). This resulted in 1756 $\pm$ 228 trials that were actually used for individual training and testing on the held out test set. The results for the hybrid set that are presented in the following sections correspond to the average of the 5 individual performances.
 
\section{Results}

\subsection{Signal quality: in-lab and in-car experiments}

To investigate the effect of the less-controlled real-life driving conditions on the quality of the recordings, we compared the data from in-lab and in-car experiments on several parameters that reflect the noise level of the EEG signals: trial rejection rates, ERP RMS and ERP PSD \cite{kuziek2017transitioning}.
The two-step trial rejection process described in the methods eliminated on average less than 10 \% of the trials from the analysis for each subject, with no significant difference between the grand average rejection rates in the two conditions (Table \ref{tab:rejection_rates}); 10.5$pm$4.94 \% in lab and 8.7$pm$5.24 \% in-car; p = 0.39, Welch's t-test). For both in-lab and in-car conditions the artifact rejections (Step 2) originated mostly from channel Fp1 and were most probably due to the eye-related movements; removing channel Fp1 from the analysis significantly decreased the percentage of total eliminated trials (Table \ref{tab:rejection_rates}; p = 0.0001, Welch's t-test).

\setlength\tabcolsep{2pt}
\begin{table}[h!]
\centering
\caption{Trials rejection rates}
 \begin{tabular}{|c||c|c|c|} 
 \hline
 \multicolumn{4}{|c|}{LAB: Trials rejection rates; median$\pm$std, \%}\\
 \hline
  & Step 1: Due to data loss & Step 2: Due to artifacts & Total\\ 
 \hline\hline
 Cz, Pz, Fp1 & 4.2$\pm$3.8 &  6.05$\pm$4.8 &  10.5$\pm$4.94 \\ 
 \hline
 Cz, Pz & 4.2$\pm$3.8 &  0.6$\pm$1.9 &  6.62$\pm$5.1\\ 
 \hline
 \end{tabular}
 \begin{tabular}{|c||c|c|c|} 
 \hline
 \multicolumn{4}{|c|}{CAR: Trials rejection rates; median$\pm$std, \%}\\
 \hline
  & Step 1: Due to data loss & Step 2: Due to artifacts & Total\\ 
 \hline\hline
 Cz, Pz, Fp1 & 0.81$\pm$0.39 &  7.8$\pm$5.32 &  8.7$\pm$5.24 \\ 
 \hline
 Cz, Pz & 0.81$\pm$0.39 &  1.18$\pm$1.47 &  2.09$\pm$1.38\\ 
 \hline
 \end{tabular}
\label{tab:rejection_rates}
\end{table}

To estimate the level of noise in individual trials we calculated average RMS scores for each trial (bandpass filtered from 0.1 to 30 Hz): first, for the baseline period (100 ms prior to stimulus onset) and second, for the whole EEG segment. RMS scores were averaged for channels Pz and Cz, then for each subject, then between subjects. The results in Fig. \ref{fig:rms_psd} A show comparable RMS values for in-lab and in-car data for both baseline and ERPs, with no significant differences in the baseline values. The ERP RMS scores in-car were statistically higher than in laboratory conditions (p = 0.01, Welch's t-test for unequal variances), indicating higher variability of the responses, which could reflect the increased mental load of real-driving scenario. 

\begin{figure}[!t]
    \centering
    \includegraphics[width=0.99\linewidth]{2019_IEEE_BioHub/figures/rms_baseline_ERP_psd_p3_long_small.jpg}
    \caption{EEG signals in in-lab and in-car conditions. A, Grand average ERP RMS values (channels Cz and Pz combined) of baseline and whole trial ERP for both conditions; individual subjects' data are shown as black dots within the boxplots. B, Grand average PSD values of channel Pz. Shaded areas show standard deviation across participants, gray area indicates statistical significant differences between in-lab and in-car conditions (p $<$ 0.05). C, Median grand averaged EEG responses for target and non-target trials; shaded area shows standard error of the means.}
    \label{fig:rms_psd}
 \end{figure}
 
Grand average PSD values were computed in range 0.1-30 Hz for in-lab and in-car conditions. The spectra showed statistically significant differences between in-lab and in-car recordings in the 3-6.3 Hz frequency window for both channels Pz and Cz (p $<$ 0.05, t-test with Bonferroni correction in the 0-20 Hz window) with no difference between channels (p $>$ 0.05, t-test with Bonferroni correction; Fig. \ref{fig:rms_psd} B shows only Pz values). Additionally, PSD spectra for the in-lab recordings had a noticeable peak around alpha frequency. 
Comparison between the grand average responses to target and non-target stimuli in laboratory and driving scenarios shows almost identical P300 ERP amplitudes and shapes for both conditions, with clear P300 ERP component appearing around 400-500 ms after stimulus onset both in channel Pz and Cz (Fig. \ref{fig:rms_psd} C, data shown for channel Pz). For both conditions and both channels target responses were statistically different from the non-target ones in the 0.4-0.5 s window (p $<$ 0.05, t-test with Bonferroni correction), wit no difference between target responses in different conditions.

\subsection{Classification results}
Table \ref{tab:classification_scores} shows the single-subject classification accuracies achieved by the different models trained on the three training sets (in-lab, hybrid, in-car) and evaluated on the last in-car session of the corresponding subject, held out as test set. All performances are above chance level, which is equal to 16 \% for the 6-class task. Regarding the performance of different classification models, 4 groups one-way ANOVA with a Turkey post hog test showed statistically significant difference only between the deep and the shallow networks (p $<$ 0.05). However, considering the mean performance, the deep network provides the best scores (53 \% mean classification accuracy), following by the random forest and the custom intermediate network (47 \% and 43 \%, respectively), and the shallow network (34 \%). There were no statistically significant differences among the classification accuracies achieved with the three different training configurations (one-way ANOVA p-values for all models were $>$ 0.05), although the hybrid version tend to perform slightly better than the other test configurations. Further experiments with increased number of participants would be required to evaluate this tendency. The results highlight a key advantage of the deep learning approach: its flexibility compared to a classical feature-based algorithm. While the classical approach requires a carefully constructed pull of relevant features, based on the extensive prior knowledge of the analysed data, one deep learning architecture could achieve state-of-the-art performance in different tasks, such as the ERP task, described in the present study and the MI task described by \cite{schirrmeister2017deep}. On the other hand, the classical ERP approach is less computationally demanding and in this particular task shows only slightly inferior performance compared to the deep network. 
\setlength\tabcolsep{2pt}
\begin{table}[h!]
\centering
\caption{Performance rates for different classification models}
 \begin{tabular}{|c||c|c|c|c|c|c|c|c|c|c|} 
  \hline
 \multicolumn{11}{|c|}{Lab}\\
 \hline
  & s001 & s002 & s003 & s004 & s005 & s006 & s007 & s008 & s009 & s010 \\ 
 \hline\hline
 Random Forest & 0.49 & 0.31 & 0.33 & 0.37 & 0.49 & 0.51 & 0.58 & 0.37 & 0.55 & 0.60 \\ 
  \hline
 Shallow Net & 0.36& 0.25 & 0.43 & 0.18 & 0.25 & 0.43 & 0.42 & 0.24 & 0.35 & 0.58 \\ 
  \hline
 Deep Net & 0.53 & 0.41 & 0.37 & 0.33 & 0.45 & 0.61 & 0.71 & 0.39 & \textbf{0.59} & 0.68 \\ 
  \hline
 Intermediate Net & 0.45 & 0.27 & 0.37 & 0.27 & 0.41 & 0.53 & 0.56 & 0.27 & 0.53 & 0.55 \\ 
 \hline
 \hline
 \end{tabular}
 \centering
 \begin{tabular}{|c||c|c|c|c|c|c|c|c|c|c|} 
  \hline
 \multicolumn{11}{|c|}{Hybrid}\\
 \hline
  & s001 & s002 & s003 & s004 & s005 & s006 & s007 & s008 & s009 & s010 \\ 
 \hline\hline
 Random Forest & 0.59 & 0.38 & 0.33 & 0.35 & 0.45 & 0.66 & 0.55 & 0.37 & 0.52 & 0.59 \\ 
  \hline
 Shallow Net & 0.45 & 0.31 & 0.43 & 0.2 & 0.27 & 0.43 & 0.44 & 0.25 & 0.33 & 0.40 \\ 
  \hline
 Deep Net & 0.55 & 0.45 & 0.35 & 0.37 & \textbf{0.53} & \textbf{0.69} & \textbf{0.69} & 0.45 & \textbf{0.59} & \textbf{0.70} \\ 
  \hline
 Intermediate Net & 0.49 & 0.41 & \textbf{0.45} & 0.29 & 0.45 & 0.53 & 0.52 & 0.33 & 0.55 & 0.57 \\ 
 \hline
  \hline
 \end{tabular}
 \centering
 \begin{tabular}{|c||c|c|c|c|c|c|c|c|c|c|} 
  \hline
 \multicolumn{11}{|c|}{Car}\\
 \hline
  & s001 & s002 & s003 & s004 & s005 & s006 & s007 & s008 & s009 & s010 \\ 
 \hline\hline
 Random Forest & 0.60 & 0.41 & 0.31 & \textbf{0.41} & 0.35 & \textbf{0.69} & 0.56 & 0.43 & 0.47 & 0.60 \\ 
  \hline
 Shallow Net & 0.45 & 0.27 & 0.31 & 0.18 & 0.24 & 0.51 & 0.42 & 0.25 & 0.35 & 0.30 \\ 
  \hline
 Deep Net & \textbf{0.62} & \textbf{0.51} & 0.35 & 0.33 & 0.37 & 0.75 & 0.65 & \textbf{0.49} & 0.51 & \textbf{0.70} \\ 
  \hline
 Intermediate Net & 0.51 & 0.35 & 0.31 & 0.24 & 0.20 & 0.55 & 0.62 & 0.24 & 0.51 & 0.57 \\ 
 \hline
 \end{tabular}
 \label{tab:classification_scores}
\end{table}

\begin{figure}[!h]
    \centering
    \includegraphics[width=0.99\linewidth]{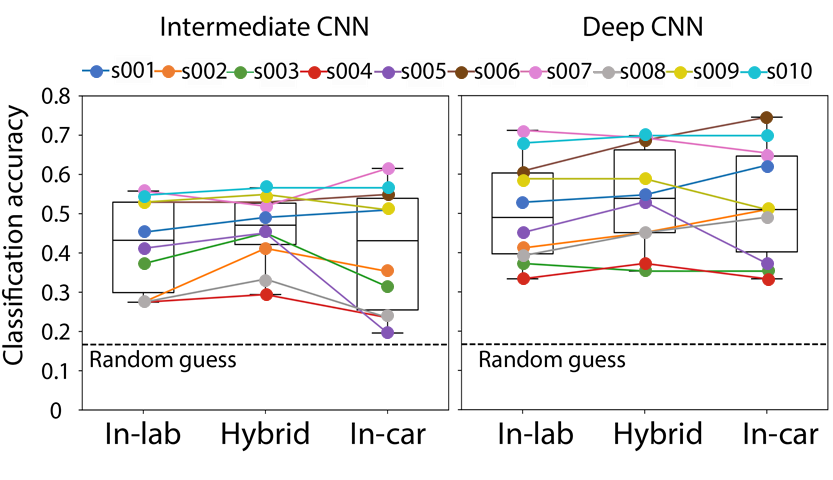}
    \caption{Subject-specific performance achieved on the test session by the intermediate and the deep networks varying the training data. In-lab and in-car models are trained only on trials recorded in lab and in car respectively, whereas hybrid models are trained on the combination of the two.}
    \label{fig:prediction_performance}
 \end{figure}
Fig. \ref{fig:prediction_performance} shows the performance of the custom intermediate CNN compared with the performance of the deep network. The fact that the single-subject models trained on exclusively laboratory (in-lab) data did not, on average, show a noticeable decrease in performance compared to the other two training scenarios is promising. It suggests that for the proposed BCI, the initial classification model could be built and pre-trained in advance in a controlled laboratory environment, without a need for collecting training data in real-life driving conditions, which are less safe and much more challenging.

 \subsection{EEG correlates}
  \begin{figure*}[!t]
    \centering
    \includegraphics[width=0.80\linewidth]{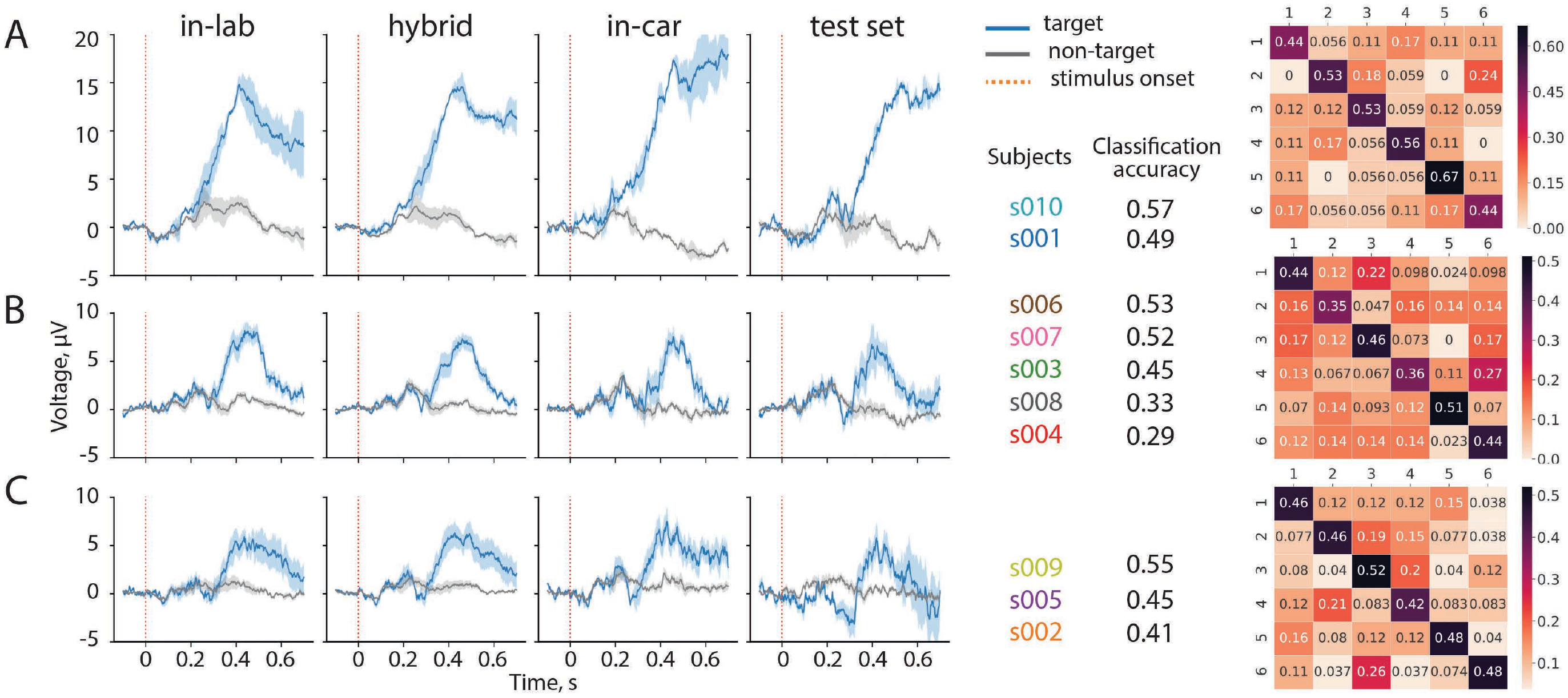}
    \caption{Grand average ERP responses for different groups. Panels A, B and C show median ERP responses to target and non-target stimuli for groups 1, 2 and 3, respectively, calculated for the training sets (in-lab, hybrid, in-car) and the test set (last in-car session). Shaded areas show standard error of the medians. The columns in the middle specify the subjects in each group (color-coded as on the Fig. \ref{fig:prediction_performance}), along with their classification performance for the hybrid model;  the plots on the right are confusion matrices for the hybrid model for the corresponding group.}
    \label{fig:global_plot}
 \end{figure*}
 To investigate the EEG features that could potentially underlie the classification results, we examined ERP responses to target and non-target stimuli for each subject and each training condition. For each subject, average peak ERP amplitudes and peak-to-peak amplitudes of target responses calculated for the three training sets were significantly different from the non-target ones (Welch's t-test, p $<$ 0.05). We then compared the peak responses between different training sets, averaging the values for all subjects to compute grand averages. The calculated values were equal to 8.7 $\pm$ 4.2 $\mu$V, 9.8 $\pm$ 5.6 $\mu$V and 8.6 $\pm$ 4.1 $\mu$V for in-lab, in-car and hybrid sets, respectively, with no statistically significant difference between the groups (Welch's t-test, p = 0.6, 0.58 and 0.96).\\
 Based on shape, quality and stability between in-lab and in-car experiments of the EEG responses, the subjects could be divided in two distinct groups. Group 1 comprised subjects s001 and s010 whose responses could not be characterized as classical ERPs (Fig. \ref{fig:global_plot}, panel A), and were most probably affected by non brain-related contamination (not identified and rejected by the threshold-based rejection approach), perhaps reflecting involuntary eye movements following the target presentation. Considering that the responses to the target stimuli in this group were particularly prominent in terms of the amplitude and duration, it was perhaps not surprising that these subjects had a high classification performance in all training scenarios.  Group 2 included subjects whose responses in both in-lab and in-car settings could be described as canonical ERPs, with an identifiable positive peak around 450 ms after the target presentation (Fig. \ref{fig:global_plot}, panel B).  Considering the oddball paradigm used for collecting the data, clear presence of this peak on centro-parietal electrodes (Cz, Pz), its characteristic shape and latency, we could assume that it includes a P300 ERP component that becomes prominent in response to the target stimulus. The main group (group 2) put together the majority of subjects: s003, s004, s006, s007 and s008, with varied classification scores, including both poorly-performing subjects (s004) and subjects that were among the best (s006, s007). Three remaining subjects (s002, s005 and s009) had ambiguous difficult to interpret responses, that could not be clearly attributed as belonging to either group 1 (responses with non brain-related contamination) or group 2 (clear ERPs) and therefore were clustered together to avoid drawing any conclusions from these data (Fig. \ref{fig:global_plot}, panel C).
 \begin{figure}[!t]
    \centering
    \includegraphics[width=0.99\linewidth]{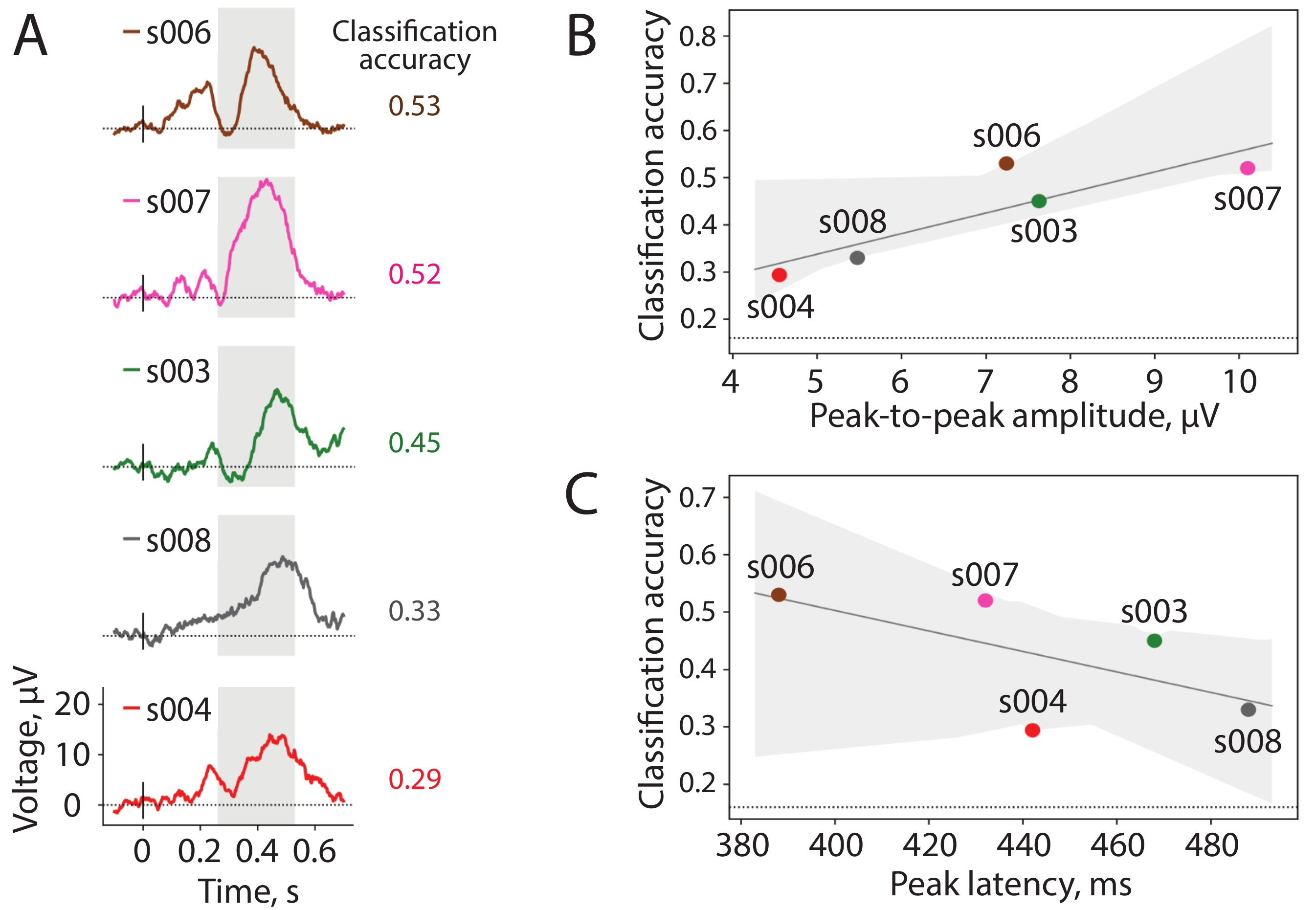}
    \caption{ERP responses of the subjects in the group 2. A, Median target responses of the hybrid training set for each subject in group 2, shown in descending order of classification accuracy. Solid vertical lines indicate stimulus onsets, dotted lines show 0 $\mu$V levels, shaded areas mark the analysed time window. Panels on the right show correlation between classification accuracy and peak-to-peak amplitude (B; $\rho$ = 0.86) and P300 peak latency (C; $\rho$ = -0.63). Subjects' data are color-coded as on the Fig.\ref{fig:prediction_performance}.}
    \label{fig:p300_amplitude_latency}
 \end{figure}
 We focused on the ERP responses of the subjects in the main group (group 2) and calculated average P300 peak amplitudes, peak-to-peak ERP amplitudes and P300 peak latencies in the time window [250:530] ms from stimulus onset for each subject and looked for the correlation between these parameters and classification accuracy for the hybrid model. Looking at the median ERP responses of individual subjects and corresponding classification scores (Fig. \ref{fig:p300_amplitude_latency}, panel A), we can see a clear trend: subjects with more distinct ERP shapes in the hybrid training set have higher performance. In particular, classification scores positively correlate with peak-to-peak amplitude (Pearson correlation coefficient $\rho$ = 0.86; panel B) and peak P300 amplitude ($\rho$ = 0.58), and negatively correlate with the peak latency ($\rho$ = -0.63; panel C). Similar correlations are found for the test set ($\rho$ = 0.85, 0.5 and -0.63, respectively). These results indicate that the learned model is likely grounding the discrimination between target and non-target on the ERP-related features of the responses, and, in particular, on the P300 component. Stronger correlation of the classification results with peak-to-peak amplitude compared to P300 amplitude or latency, suggests that other ERP components are presumably picked up and taken into account by the CNN, playing an important role in the discrimination between target and non-target EEG responses. Considering that none of the aforementioned correlations are perfect, other EEG features could be affecting classification results, identification of which requires further investigation.
 
\section{Limitations}
The present study represents a first evaluation of the usage of a portable EEG-based BCI in real-life driving scenario to control a car infotainment menu. Authors acknowledge that the current state of the system still has limitations that are planned to be addressed in the continuation of the study and could greatly improve the overall performance and reliability of the whole device. One drawback of the present BCI is a relatively high ratio of rejected trials, originating from two main sources: loss of data during the wireless transmission and the artifacts present in the recorded EEG signals (Table \ref{tab:rejection_rates}). The first issue could be solved by improving the wireless connection or switching to a direct data transmission, and would eliminate a significant portion of rejections. 
In the present state of the system, the majority of rejections are due to channel Fp1 and most probably reflect eye-related artifacts. To mitigate this issue we plan to implement one of the available algorithms for ocular artifact removal and integrate an eye-tracker to ease the detection of eye-related artifacts in the training phase and confirm the operation of the BCI with only peripheral vision.
Due to the differences between the experimental protocols in the in-lab and in-car settings, the focus of this study was not the direct comparison of the EEG brain responses between conditions but the applicability of the P300 paradigm to control the infotainment menu in a real driving scenario. 
Nevertheless, a preliminary analysis showed that the average ERP responses obtained in laboratory and driving experiments were very similar and indeed we were able to use both to reliably classify between target and non-target stimuli with comparable prediction scores. 
\section{Conclusions}
This study demonstrates the practical usability of a wearable EEG-based BCI in a real driving scenario, in particular, for solving the problem of item selection from the car infotainment menu while driving. Notably, such a system does not require any manual selection, and achieves its objective via a robust yet simple infrastructure, implementing  an oddball-like stimulus presentation paradigm.
Overall, the system achieved a promising performance with all the evaluated subjects despite different features characterizing their ERP responses. The results of a classical BCI classification approach, based on manual feature extraction, confirm that the P300-related features are the most relevant for the discrimination of target and non-target responses.\\ 
Results suggest a two-step training as an ideal procedure for real driving conditions: the core of the subject-specific classification models should be trained on laboratory data alone, while a continuous refinement while driving should be effective in improving and stabilizing the performance.
This would also help tackling the non-stationarity of the EEG signals.
Nevertheless, further investigations are foreseen to improve the system performance, in particular in terms of classification accuracy and number of required repetitions. 


\section*{Acknowledgments}
The authors gratefully acknowledge the European Commission for its support of the Marie Sklodowska Curie program through the H2020 ETN PBNv2 project (GA 721615) as well as Toyota Motor Europe for providing the infrastructure needed for this experiment. We would also like to thank the participants for their collaboration.



%

\bibliography{root} 
\bibliographystyle{ieeetr}

\end{document}